\documentclass[research]{fcs}
\usepackage{multirow}
\usepackage{makecell}
\usepackage{afterpage}
\usepackage{graphicx}
\usepackage{subfigure}

\usepackage{changes}

\title{ROS package search for robot software development: a knowledge graph-based approach}

\author[1,2]{Shuo WANG}
\author*[1,2]{Xinjun MAO}
\author[3]{Shuo YANG}
\author[4]{Menghan WU}
\author[1,2]{Zhang ZHANG}
\address[1]{College of Computer, National University of Defense Technology, Changsha 410000, China}
\address[2]{Key Laboratory of Software Engineering for Complex Systems, Changsha 410000, China}
\address[3]{College of Systems Engineering, National University of Defense Technology, Changsha 410000, China}
\address[4]{College of Computer Science and Technology, Zhejiang University, Hangzhou 310058, China}

\fcssetup{
  received       = {month dd, yyyy},
  accepted       = {month dd, yyyy},
  corr-email     = {xjmao@nudt.edu.cn},
}
\begin{abstract}
In recent years, ROS (Robot Operating System) packages have become increasingly popular as a type of software artifact that can be effectively reused in robotic software development. Indeed, finding suitable ROS packages that closely match the software's functional requirements from the vast number of available packages is a nontrivial task using current search methods. 
The traditional search methods for ROS packages often involve inputting keywords related to robotic tasks into general-purpose search engines  (e.g., Google) or code hosting platforms  (e.g., Github) to obtain approximate results of all potentially suitable ROS packages. 
However, the accuracy of these search methods remains relatively low because the task-related keywords may not precisely match the functionalities offered by the ROS packages. 
To improve the search accuracy of ROS packages, this paper presents a novel semantic-based search approach that relies on the semantic-level ROS Package Knowledge Graph  (RPKG) to automatically retrieve the most suitable ROS packages. 
Firstly, to construct the RPKG, we employ multi-dimensional feature extraction techniques to extract semantic concepts, including \emph{code file name}, \emph{category}, \emph{hardware device}, \emph{characteristics}, and \emph{function}, from the dataset of ROS package text descriptions. 
The semantic features extracted from this process result in a substantial number of entities (32,294) and relationships (54,698). 
Subsequently, we create a robot domain-specific small corpus and further fine-tune a pre-trained language model, BERT-ROS, to generate embeddings that effectively represent the semantics of the extracted features. These embeddings play a crucial role in facilitating semantic-level understanding and comparisons during the ROS package search process within the RPKG. 
Secondly, we introduce a novel semantic matching-based search algorithm that incorporates the weighted similarities of multiple features from user search queries, which searches out more accurate ROS packages than the traditional keyword search method. 
To validate the enhanced accuracy of ROS package searching,
we conduct comparative case studies between our semantic-based search approach and four baseline search approaches: ROS Index, GitHub, Google, and ChatGPT. The experiment results dem-onstrate that our approach achieves higher accuracy in terms of ROS package searching, outperforming the other approaches by at least 21\% from 5 levels, including top1, top5, top10, top15, and top20.

\end{abstract}

\begin{document}

\section{Introduction}


As autonomous robots are widely used in human society, developing robotic software has become a prevalent practice in the robotic community, responding to the high demand from robot consumers. However, the development of robotic software has always been challenging due to its requirement of interdisciplinary knowledge in AI, control, and software engineering, as well as significant programming effort for implementing robotic software components \cite{kolak2020takes}. ROS has emerged as the de facto standard for robotic software development frameworks. It offers a pleth-ora of reusable ROS packages, each implementing an independent robotic software functionality such as task planning, actuation, and visual sensing \cite{quigley2009ros}. By reusing and combining these packages, developers can quickly construct a robust robotic software framework and implement task-specific algorithms when needed \cite{pichler2019can,cousins2010sharing}.

As of August 10, 2023, there are 2,644 software repositories and 7,570 packages that have been developed using the ROS framework\footnote{\url{https://index.ros.org/stats/}}. However, searching for the most suitable ROS package for a specific robotic software functionality remains a time-consuming and often inaccurate process. Existing search methods primarily rely on direct text matching, where a set of task-related keywords are inputted into a search engine to obtain approximate results. Unfortunately, this approach often yields irrelevant ROS packages due to its disregard for the semantic information of the tasks, which considerably prolongs the searching process. 

Currently, four mainstream approaches are commonly used to search for ROS packages. (1) Searching using the ROS Index website, which serves as a gateway to search for ROS and ROS2 resources, including packages, repositories, system dependencies, and documentation\footnote{\url{https://index.ros.org/}}. It allows developers to enter keywords and phrases in the search bar to find suitable packages, ranked by relevance calculated using the BM25 algorithm \cite{robertson2009probabilistic}. However, ROS Index often outputs irrelevant ROS packages due to its reliance on syntactic word matching. In addition, it lacks semantic comprehension, making it challenging to deal with synonyms and abbreviations. (2) Searching GitHub repositories, where most ROS packages can be found. GitHub matches user queries with names and descriptions of all repositories, yielding search results based on character syntactic matching \cite{booth2016github}. For instance, the word ``do" may be matched to ``docker" in the description of a repository. Since the search repositories must include all the words from the user queries, it is not uncommon for GitHub to return no results for queries with up to three words. Furthermore, each repository often contains more than one ROS package. (3) Searching using the Google search engine. By utilizing a custom search combining ROS Index and GitHub, developers can limit their search scope by adding ``site:ros.org OR site:github.com" to their queries. All three search methods rely on text syntactic matching without extracting and understanding the semantics of ROS packages, leading to mismatches between search results and user queries.  (4) Searching using ChatGPT. Recent advancements in Large Language Models (LLMs) have made ChatGPT\footnote{\url{https://chat.openai.com/}} one of the most powerful semantic models. ChatGPT improves the understanding of the semantics of ROS packages, allowing users to interact in natural language to find ROS packages. However, ChatGPT may struggle with domain-specific words such as ``TurtleBot2" and ``TurtleBot3", limiting its accuracy in searching ROS packages.


These four search approaches for ROS packages only utilize the text description of ROS packages syntactically and do not exploit the rich features of ROS packages on multiple dimensions. To gain a comprehensive understanding and representation of ROS packages, we require a high-level abstraction of ROS packages to identify the multi-dimensi-onal concepts and relationships within them to facilitate the search task. Knowledge graphs have been widely applied in various fields such as software engineering in recent years as they provide structured knowledge representation and facilitate knowledge retrieval. To represent the multi-dimens-ional concepts and relationships of ROS packages in a structured format and support the search task, we propose using a knowledge graph as a domain knowledge base for ROS packages. Additionally, keyword matching-based search methods are not suitable for knowledge graphs, necessitating the design of a new knowledge graph-based search algorithm for ROS packages.

To improve the search accuracy of ROS packages, this paper presents a novel semantic-based search approach that relies on the semantic-level ROS Package Knowledge Graph  (RPKG) to automatically retrieve the most suitable ROS packages. Firstly, to enrich the semantics of ROS packages, we utilize multi-dimensional feature extraction techniques to extract the semantic concepts of categories, functions, characteristics, associated robot hardware, etc., based on the dataset of ROS package text descriptions. RPKG is constructed based on the extracted semantic concepts of 32,294 entities  (11 types) and 54,698 relationships  (10 types), which supports retrieving knowledge on ROS packages in a structured way. To represent the semantics of ROS package features, we build a corpus in the robot domain and further fine-tune the pre-trained language model BERT, which improves the semantic similarity matching for natural language. Secondly, to support the search for ROS packages, we propose a novel semantic matching-based search algorithm by incorporating weighted similarities of multiple features of the user’s search queries. To assess the quality of extracted semantic features from ROS packages, we engage two robot software developers to evaluate the accuracy of sampled features across multiple dimensions, which yields an accuracy rate of over 91.6\%. To validate the enhanced accuracy of ROS package searching, we conduct comparative case studies, comparing our semantic-based search approach with four baseline search approaches: ROS Index, GitHub, Google, and ChatGPT. The experimental results demonstrate that our approach achieves higher accuracy in terms of ROS package searching, outperforming the other approaches by at least 21\% from 5 levels, including top1, top5, top10, top15 and top20.

To summarize, the main contributions of this paper are as follows:

\begin{itemize}
\item To understand and represent ROS packages more comprehensively, we first establish a high-level abstraction view of ROS packages, which discusses the main concepts of each ROS package and the main relationships between ROS packages.
\item To enrich the semantics of ROS packages, we propose a multi-dimensional feature extraction technique and construct a ROS package knowledge graph, which more comprehensively represents ROS packages from multiple dimensions in a structured way.
\item To support the knowledge graph-based search for ROS packages, we design a novel ROS package feature fusion matching algorithm by considering weighted similarities of multiple features from users' queries, which outperforms existing methods by at least 21\% in search accuracy.
\end{itemize}

The remainder of this paper is organized as follows. We discuss the current related works in Section \ref{section:related-work}. Section \ref{section:abstraction} proposes the high-level abstraction view of ROS packages. Section \ref{section:rpkg-construction} introduces the construction of RPKG. Section \ref{section:search} presents the RPKG-based ROS search approach. Evaluations are shown in Section \ref{section:evaluation}. We close with our conclusions and future work in Section \ref{section:conclusion}.

\section{Related work}
\label{section:related-work}

\subsection{Code Reuse in ROS}
In recent studies, researchers pay more attention to the ecosystem of ROS and the dependencies between ROS packages, among other factors. Estefo et al. conducted an empirical study consisting of interviews and a survey with ROS developers to study the evolution challenges faced by the ROS ecosystem, especially the issues surrounding package reuse and bottlenecks to contribute to existing packages \cite{estefo2019robot}. They also studied the code quality of the ROS ecosystem through an analysis of code duplication in ROS launchfiles \cite{estefo2015code}. Jiang and Mao conducted empirical research on questions about using launch files on Stack Overflow and ROS Answers (a Q\&A platform for ROS) \cite{jiang2022exploring}. Witte and Tichy analyzed ROS nodes and their launch files to check consistency and issue warnings for robot software in ROS \cite{witte2018checking}. Malavolta et al. presented an observational study that unveils the state-of-the-practice for architecting ROS-based systems and provides guidelines for roboticists to properly architect their ROS-based systems \cite{malavolta2021mining}. Kolak et al. studied the growth of the ROS ecosystem and the collaboration among its members \cite{kolak2020takes}. They also analyzed the dependencies between packages to study the structure of the ROS ecosystem. Canelas et al. reported the challenges and experiences of learning and using ROS from the perspective of ROS newcomers. They found that there are still many common misunderstandings and mistakes when newcomers use ROS to develop their robotic software \cite{canelas2022experience}. Macenski et al. reviewed the philosophical and architectural changes of ROS 2, which powers the robotics revolution \cite{macenski2022robot}. The aforementioned studies mainly focus on the usage and development of ROS but do not specifically investigate the problem of finding suitable ROS packages for robot software.

\subsection{Knowledge Graph in Software Engineering}

Recently, there has been increasing interest in applying knowledge graphs in software engineering due to their powerful ability in knowledge representation, reasoning, and support for knowledge se-arch \cite{sun2019know, bo2022towards}. However, very few studies have focused on the application of knowledge graphs in robotics. Tenorth and Beetz designed the knowledge processing system KnowRob to equip robots with the capability to organize information in reusable knowledge chunks and to perform reasoning in an expressive logic \cite{tenorth2013knowrob}. Sun et al. constructed a ROS message knowledge graph (RMKG) to assist users in finding appropriate ROS message types \cite{bo2022towards}. They mainly extracted features of ROS messages from their names and proposed a fuzzy matching algorithm to support the search for ROS messages. Jiamou et al. developed open information extraction (OpenIE) techniques to extract candidates for task activities, activity attributes, and activity relationships from Android programming task tutorials \cite{sun2019know}. Based on these extracted features, they built a knowledge graph TaskKG, which enables activity-centric knowledge search. Zou et al. proposed a code-centric software knowledge model and provided a two-layer plugin framework for knowledge graph construction and software Q\&A \cite{zou2021software}.

\subsection{Search Methods in Software Development}
Nowadays, developing software from scratch is still quite difficult and time-consuming. Software reuse has proven to be an advantageous method for simplifying software development. However, before reusing software or a component, it is necessary to find a suitable one that meets the software requirements. In the past decades, researchers have conducted numerous studies on software search and recommendation. Seacord et al. described Agora as a prototype search engine that automatically generates and indexes a worldwide database of software products \cite{seacord1998agora}. López Pino B E developed a tool to retrieve data from GitHub and ROS Answers, and then proposed a recommender algorithm to answer questions on ROS Answers \cite{lopez2019sistema}. Agora allows users to query Java components with methods or properties as inputs. Yang and Nyberg investigated two problems: search task suggestion using procedural knowledge and automatic procedural knowledge base construction from search activities \cite{yang2015leveraging}. They proposed the creation of a three-way parallel corpus of queries, query contexts, and task descriptions, and reduced both problems to sequence labeling tasks. They also proposed a supervised framework to identify queryable phrases from task descriptions and extract procedural knowledge from search queries and relevant text snippets. Silva et al. proposed CROKAGE (CrowdKnowledge Answer Generator), a tool that takes the description of a programming task (the query) as input and delivers relevant code examples for the task \cite{da2020crokage}. CROKAGE leverages the knowledge stored in Stack Overflow to generate solutions containing source code and explanations for programming tasks written in natural language.

\section{High-level Abstraction of ROS Packages}
\label{section:abstraction}
ROS packages, as commonly known by most roboticists, are implementation-level software artifacts that are frequently reused and programmed in various robotic software projects. In this paper, we establish a high-level abstraction view of ROS packages, which discusses the main concepts of each ROS package and the main relationships between ROS packages.

Properties related to ROS packages is various, including authors, emails, development status, last updated date, code files, descriptions, etc. However, some of these properties are unimportant or irrelevant when searching for a ROS package, so there is no need to include them in the RPKG model. We first analyze the main concepts of ROS packages and then discuss the interconnected relationships between ROS packages.

\subsection{ROS Package Concepts}

We thoroughly investigate the features present in developers' requirements to guide the search for ROS packages. For this purpose, we examine the 100 most recent questions from ROS Answers, where users have requested assistance in searching for ROS packages. We delve into each question and its corresponding ROS package answers, ultimately identifying five key concepts that significantly contribute to finding the most suitable ROS packages:
\begin{itemize}
    \item \textbf{Robot}: It specifies the robot that users use. Some ROS packages have specific scopes and adaptations for certain robots.
    \item \textbf{Sensor}: It specifies the sensor that users use.
    \item \textbf{Package category}: It specifies the category of the ROS package, such as Message package, Description package, Meta package, or Function package. The Message package is used for defining ROS messages transferred between packages. The Description package is used for describing the physical models of the robot hardware. The Meta package\footnote{\url{https://wiki.ros.org/Metapackages}} is used as a virtual package, which does not provide any functional code files, but rather references one or more related packages that are loosely grouped together. We define the Function package as a package that implements specific functions by processing data or interacting with the physical environment.
    \item \textbf{Package function}: It specifies the function that a package can provide, such as mapping, achieving navigation, or bringing up the robot.
    \item \textbf{Package characteristics}: It covers other constraints mentioned in users' questions, such as specific algorithms, environment requirements, etc.
\end{itemize}

\begin{figure}[!ht]
\centering
\includegraphics[width=1.0\linewidth]{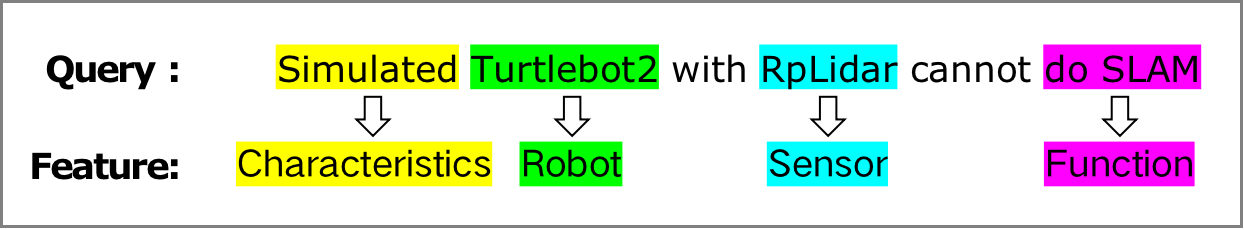}
\caption{A query sample from ROS Answers  (\url{https://answers.ros.org/question/404381/simulated-turtlebot2-with-rplidar-cannot-do-slam/})}
\label{fig:query-sample}
\end{figure}

Fig. \ref{fig:query-sample} shows a query sample from ROS Answers. ``Turtlebot2" represents the robot used; ``Rplidar" refers to the sensor of the Rplidar series; ``do SLAM" is the function of the required ROS package; and ``Simulated" is the characteristic that limits this task to a simulated environment.

\begin{figure*}[!ht]
\centering
\includegraphics[width=0.8\linewidth]{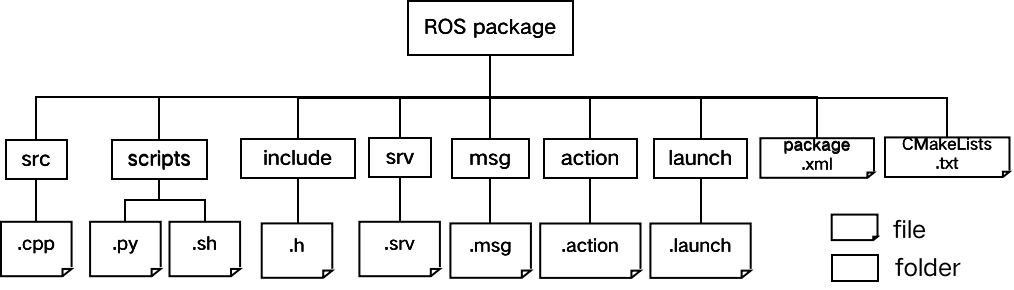}
\caption{Filesystem of the ROS package}
\label{fig:filesystem}
\end{figure*}

In addition to the previously mentioned five key concepts, we have discovered that code files can also aid in locating a ROS package in certain cases. We study the code files of ROS packages to help search for suitable ROS packages. Fig. \ref{fig:filesystem} demonstrates the file structure of a typical ROS package. In terms of the filesystem, a ROS package must contain a catkin compliant package.xml file, which provides meta-information about the package, and a CMakeLists.txt file which uses catkin. Codes of ROS nodes are often written in C++ or Python and are stored in separate /src and /scripts folders. Shell scripts are also placed in the /scripts folder. Header files for C++ programs are placed in the /include folder. The /srv and /msg folders contain service files in .srv and message files in .msg format, respectively. The /action folder contains action files in .action format. Additionally, launch files with a .launch extension are used as configuration parameters of ROS nodes.

\begin{figure}[!ht]
\centering
\includegraphics[width=1.0\linewidth]{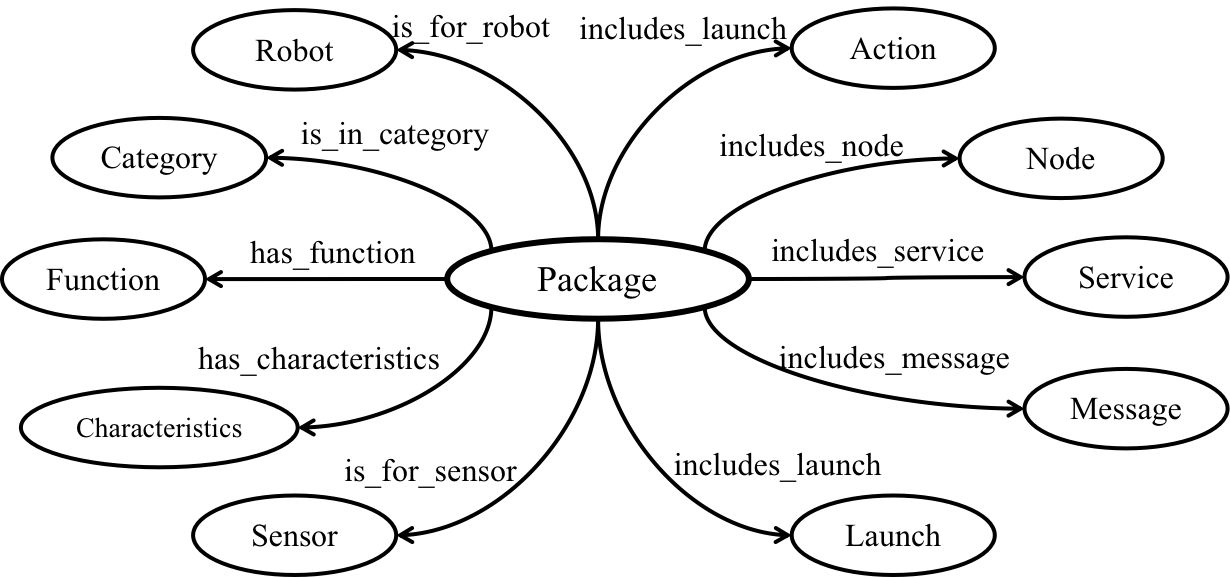}
\caption{Conceptual Schema of ROS Package Knowledge Graph}
\label{fig:model}
\end{figure}

The modeled schema, depicting the abstraction of 11 entities, is illustrated in Fig. \ref{fig:model}. Concepts related to code files include action, node, service, message, and launch.

\subsection{ROS Package Relationships}Alongside the entities, we have identified 11 key relations between ROS packages and other entities. However, for the purposes of this paper, we choose to ignore the dependency relation between packages, as it mainly contributes to software configuration.
\begin{itemize}
    \item \textbf{is\_for\_device}: Some packages have limitations in hardware devices. Package A may only be applied to Robot R.
    \item \textbf{is\_in\_category}: Different packages have different categories, such as meta package, message package, or description package. There are also differences between them in tasks, such as visualization, navigation, or manipulation.
    \item \textbf{has\_function}: A package has its own functions, such as implementing a mapping algorithm or driving a laser.
    \item \textbf{has\_characteristics}: Each package has some characteristics.
    \item \textbf{includes\_\$\{file type\}}: Generally, a ROS package is composed of nodes, services, messages, launch files, a Cmakelist file, and an XML file. As the Cmakelist file and package XML file have fixed names, they are useless for retrieving ROS packages and are removed from RPKG.
 \end{itemize}

\section{RPKG Construction}
\label{section:rpkg-construction}

\begin{figure*}[!ht]
\centering
\includegraphics[width=1.0\linewidth]{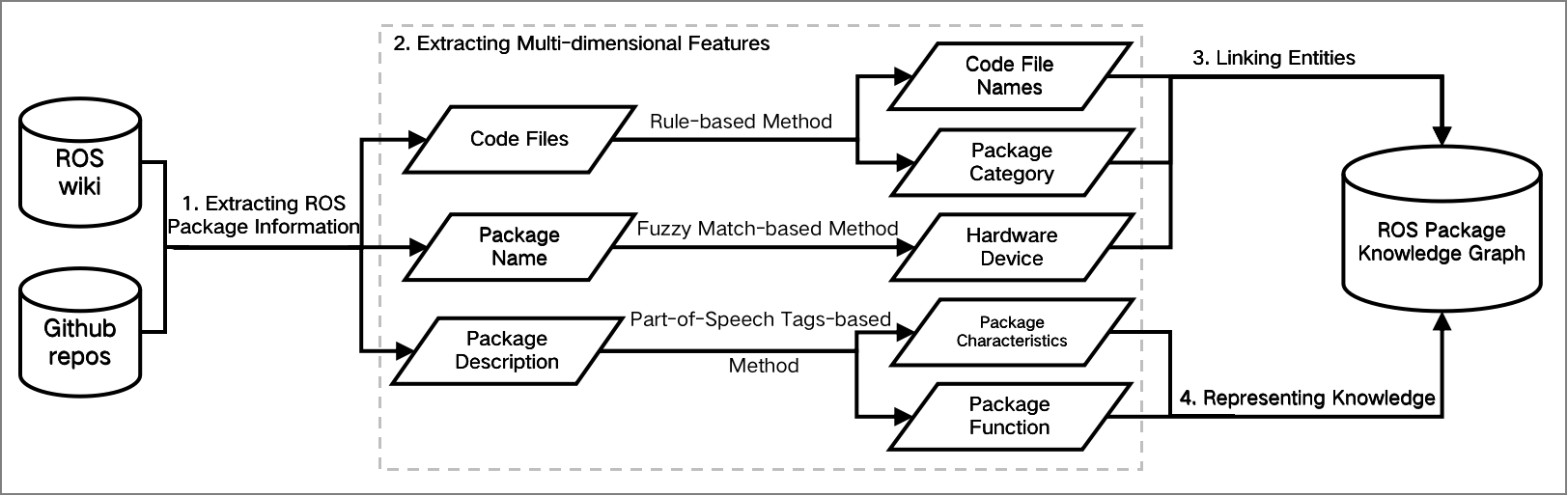}
\caption{The overview of construction approach for RPKG}
\label{fig:construction}
\end{figure*}

Fig. \ref{fig:construction} presents four key steps for constructing RPKG from ROS wiki and GitHub repositories. We first design a conceptual schema to represent the entities and relations required to search ROS packages. Then we use the popular web crawler tool Scrapy4 to crawl ROS wiki tutorials and GitHub repositories. Next, we parse each package to retrieve its code files, package name, and package description. We propose a multi-dimensional feature extraction technique to process the following: 
\begin{itemize}
    \item \textbf{Code Files}: Extracted with a rule-based met-hod to retrieve package category and names of nodes, messages, launch files, actions, and services.
    \item \textbf{Package Name}: Extracted with a fuzzy match method to retrieve the hardware device.
    \item \textbf{Package Description}: Extracted with a dependency syntactic parsing method to retrieve its function and characteristics.
\end{itemize}

In order to facilitate the search of ROS software packages, we adopt different knowledge representations for different features. For extracted features on code file names, package categories, and hardware devices, they can be easily retrieved by names. So we link them by names to entities in the ROS package knowledge graph and create corresponding relationships according to the conceptual schema in Fig. \ref{fig:model}. For features on package characteristics and package function, they are usually phrases in natural language and their semantics are difficult to retrieve directly from text. Hence, we further fine-tune a BERT model to represent knowledge of these features. Finally, using the features extracted earlier, we construct the RPKG.

\begin{figure*}[h]
\centering
\includegraphics[width=0.7\linewidth]{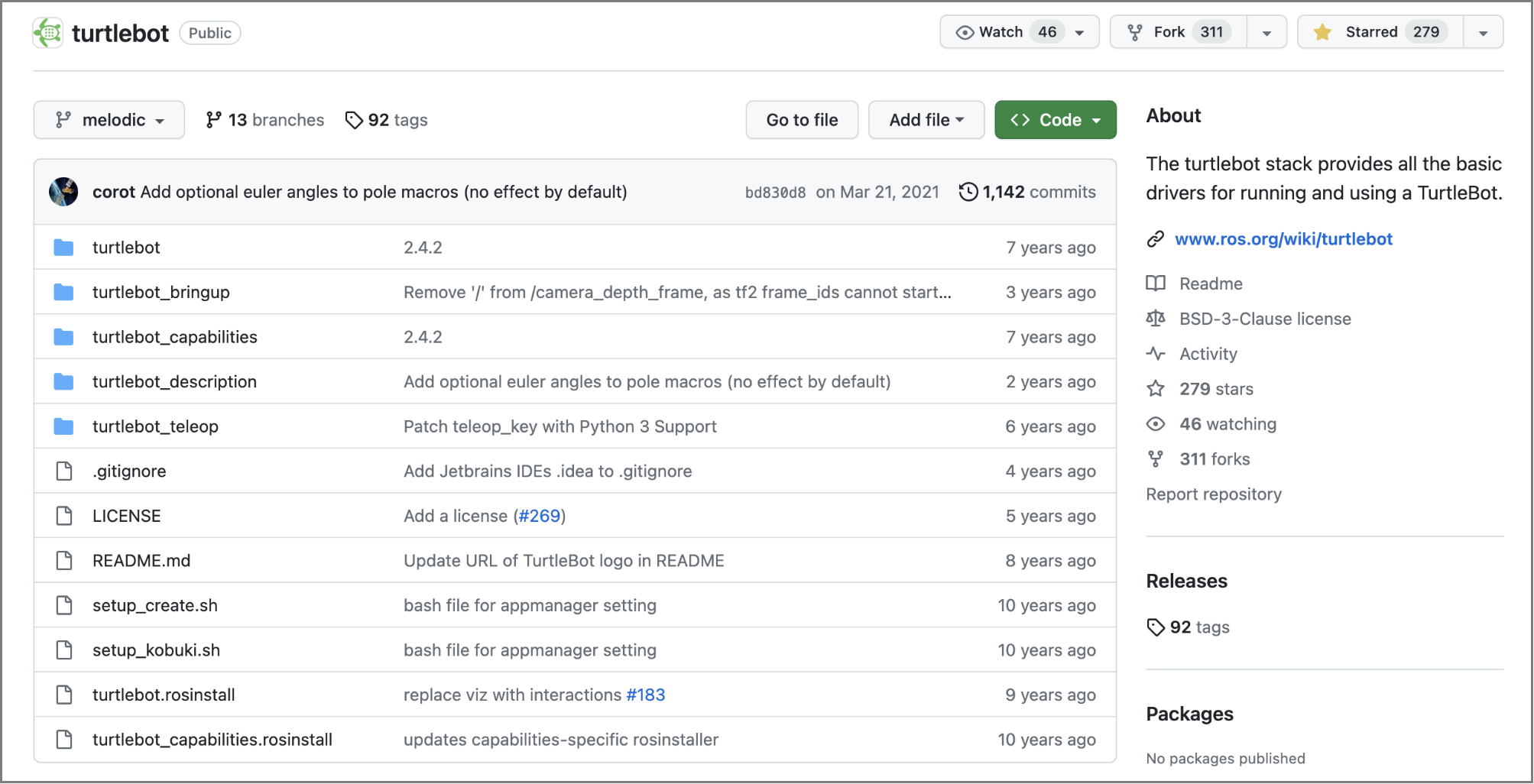}
\caption{The ``turtlebot" repository in GitHub (\url{https://github.com/turtlebot/turtlebot})}
\label{fig:turtlebot-repo}
\end{figure*}

\subsection{Extracting ROS Package Information}

\begin{figure*}[h]
\centering
\includegraphics[width=0.7\linewidth]{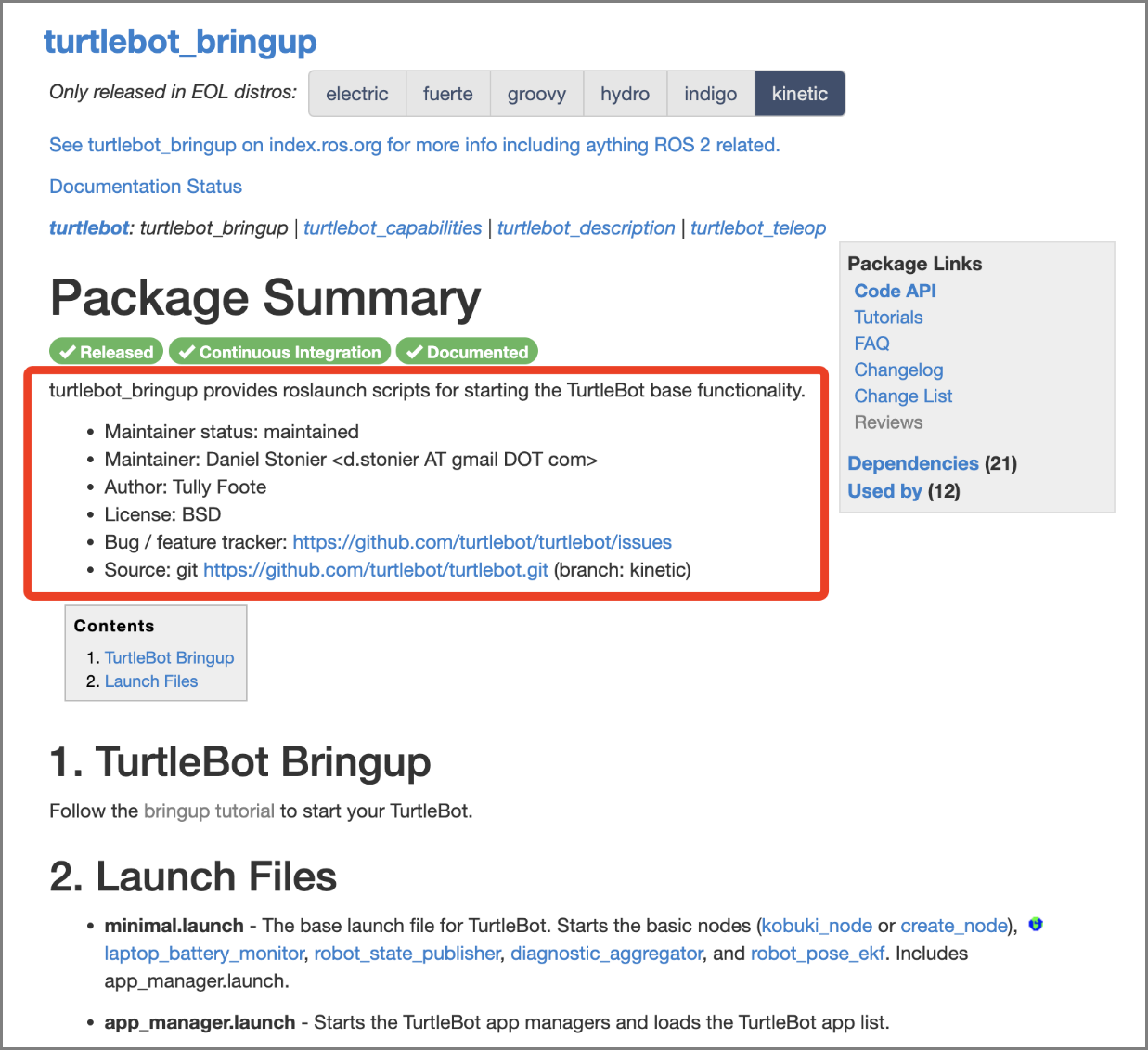}
\caption{The wiki page for the ROS package ``turtlebot\_bringup" (\url{https://wiki.ros.org/turtlebot\_bringup})}
\label{fig:roswiki}
\end{figure*}

As the ROS ecosystem continues to develop, a wealth of knowledge about ROS packages has emerged across various sources, including research papers, books, courses, and online websites. However, the collection of knowledge from these distributed sources poses a significant challenge. The ROS organization opens source a GitHub project ``rosdistro"\footnote{\url{https://github.com/ros/rosdistro}} to maintain a list of repositories for each ROS distribution. From ``rosdistro", we can retrieve the package names, repository URLs, and  all official ROS packages included in each repository, which can help us locate the code files of ROS packages on GitHub. Fig. \ref{fig:turtlebot-repo} shows the ``turtlebot" repository in GitHub. The ``turtlebot" repository includes 5 ROS packages related to the TurtleBot2 robot.

To maintain the documentation of ROS packages, the ROS organization applies the wiki system to encourage all developers to contribute their ROS packages to the ROS official wiki website \url{https://wiki.ros.org}. Fig. \ref{fig:roswiki} shows the wiki page for the ROS package ``turtlebot\_bringup". In the red box, we can see the package description ``turtlebot\_bringup provides roslaunch scripts for starting the TurtleBot base functionality.". It also provides information about the package's maintainer status, maintainer, author, license, bug or feature tracker, and source URL. Package links on the right of the red box show their dependency relations with other ROS packages. Below the red box are tutorials and launch files for the package, which may be outdated knowledge as the wiki page is not synchronized with the code repository in real-time.

After parsing the ROS packages from ROS wiki tutorials and GitHub repositories, we retrieve the code files, package names, and package descriptions for each ROS package.

\subsection{Extracting Multi-dimensional Features}

\label{section:feature}

Since the knowledge regarding the categories and contents of ROS packages follows a structured format, we can apply rules to extract these features, including category, node, service, message, launch, and action. However, when dealing with features related to robot and sensor hardware, which often have various abbreviations, direct matching becomes challenging. To address this, we employ a fuzzy match-based method to enhance the extraction of hardware features from the package name. Package descriptions typically provide information about the function and characteristics of packages in natural language, which is unstructured. To extract relevant features from these descriptions, we utilize part-of-speech  (POS) tags to identify and extract relevant phrases as features on function and characteristics.
\subsubsection{Rule-based Method}

Packages differ from each other in file structures. So we can extract the package category by applying the following four rules:
\begin{itemize}
    \item \textbf{Rule 1:} If a package only includes basic package.xml and CMakeLists.txt without any ROS node related files, then it is a meta package.
    \item \textbf{Rule 2:} If a package includes /meshes or /robots folder to describe a robot model, then it is a description package.
    \item \textbf{Rule 3:} If a package includes /msg folder to declare ROS messages, then it is a message package.
    \item \textbf{Rule 4:} If the previous three rules  are not met, then the package is a function package. 
\end{itemize}

For code files in each package, we can extract code file names by applying the following four rules:
\begin{itemize}
    \item \textbf{Rule 5:} If a file is in the ``/scripts" folder and its name ends with ``.py", then it is a Node file.
    \item \textbf{Rule 6:} If a file is in the ``/srv" folder and its name ends with ``.srv", then it is a Service file.
    \item \textbf{Rule 7:} If a file is in the ``/msg" folder and its name ends with ``.msg", then it is a Message file.
    \item \textbf{Rule 8:} If a file is in the ``/action" folder and its name ends with ``.action", then it is an Action file.
\end{itemize}

A ROS node can be compiled from C++ code files ending with ``.cpp" with ``CMakeLists.txt". So we can apply regular expressions to retrieve names of ROS nodes from ``CMakeLists.txt".

With the rule-based method, we can retrieve package categories and code file names as features of ROS packages.

\subsubsection{Fuzzy Match-based Method}

Each package name uniquely refers to a ROS package. The name of a ROS package is important as it contains key features of the package. For example, we can determine from the package name ``turtlebot3\_msgs" that it is specific to the Turtlebot3 robot and it is a ROS message package, indicating that it is used for defining message types for the Turtlebot3 robot.

\begin{figure*}[ht]
\centering
\includegraphics[width=1.0\linewidth]{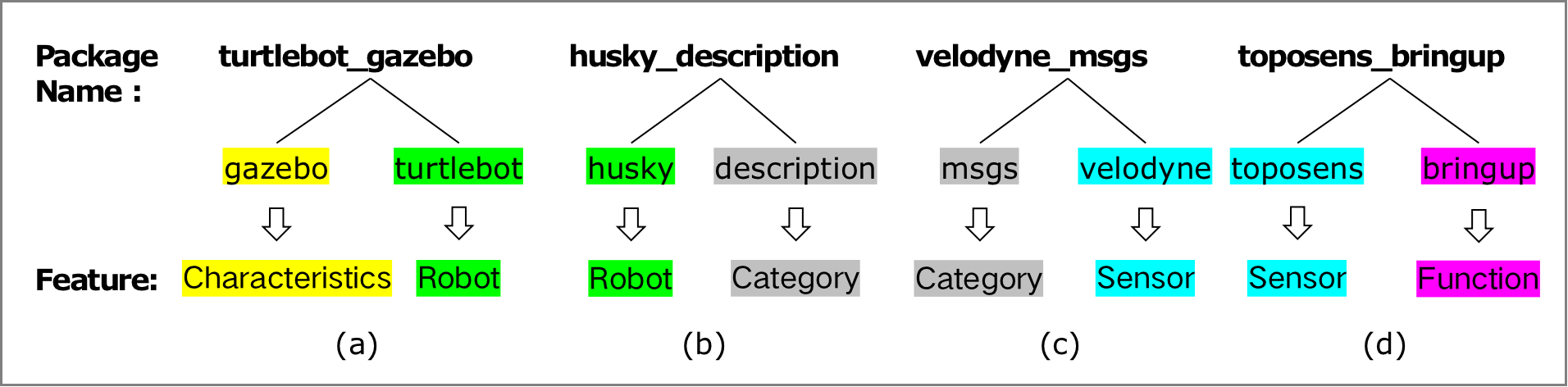}
\caption{Naming structures of 4 ROS packages}
\label{fig:package-name}
\end{figure*}

In 2015, the ROS organization provided guidelines and rules for package naming (ROS REP 0144)\footnote{\url{https://ros.org/reps/rep-0144.html}}, which also assist in extracting features from package names. Essentially, each package name uses ``\_" separators among feature words. If the package is specialized for a hardware device, the word before the first ``\_" separator usually refers to that device. For example, Fig. \ref{fig:package-name} shows the naming structure of four ROS packages. In the case of the ``turtlebot\_gazebo" package, ``turtlebot" indicates that it is suitable for the Turtlebot2 robot and ``gazebo" indicates that it works in the Gazebo simulator. ``velodyne\_msgs" provides ROS message definitions for Velodyne (a 3D lidar sensor). The ``msgs" part signifies that it belongs to the category of message packages. ``toposens\_bringup" is used for bringing up the Toposens ultrasonic 3D sensor. Here, ``bringup" refers to the function feature of this package.

To determine the related hardware device, we need to analyze the first word in the package name. However, this analysis is challenging due to the absence of an organized list of existing hardware devices in robotics. Moreover, abbreviations are common in package naming, and package developers have different naming styles. In previous works \cite{bo2022towards}, a fuzzy match-based method was proven effective in dealing with abbreviations. In this study, we apply the fuzzy match-based method to recognize the entities of robots and sensors. Firstly, we collect existing known robots and sensors as a vocabulary for fuzzy matching. We extract features of known robots from official ROS website for robots \url{https://robots.ros.org/}, which includes the robot name, robot category, website URL, wiki page URL, robot description, and related tags. Sensor names and related packages can be retrieved from ROS wiki for sensors \url{https://wiki.ros.org/Sensors}. Then, we compute a sorted fuzzy similarity score list between all hardware names and the first word in the package name. If the highest score exceeds the threshold, then we consider the corresponding hardware as the related hardware entity for the package.

To implement the fuzzy match, we utilize the Levenshtein Distance algorithm to calculate the similarity score between two words. The threshold value has an impact on the accuracy of recognizing robot and sensor features. If the threshold is set too high, many positive feature words may be missed in the fuzzy match process, resulting in a low precision rate. Conversely, if the threshold is set too low, some incorrect words may be erroneously recognized as hardware features, leading to a low recall rate. Through testing, we determine the threshold of 90 to achieve the best F1 score, striking a balance between precision and recall rates.

The fuzzy match-based extraction method consists of four steps:
\begin{itemize}
    \item \textbf{Step 1}: Split the ROS package name to obtain the first word.
    \item \textbf{Step 2}: Calculate the similarity scores of the word within the hardware vocabulary.
    \item \textbf{Step 3}: Sort the similarity scores in descending order and identify the hardware name with the highest score.
    \item \textbf{Step 4}: Add the hardware feature to the ROS package if its score exceeds the threshold of 90.
\end{itemize}

\begin{figure*}[ht]
\centering
\includegraphics[width=1.0\linewidth]{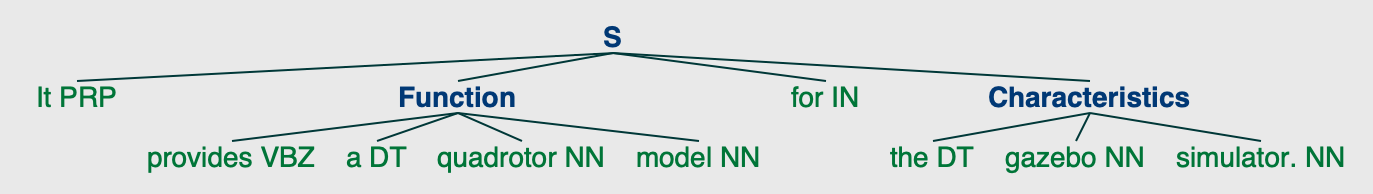}
\caption{Conceptual Schema of ROS Package Knowledge Graph}
\label{fig:pos-tag}
\end{figure*}

\subsubsection{Part-of-Speech Tags-based Method}
As the package descriptions are unstructured text, it is challenging to directly extract features from them. Inspired by the TaskKG work of Sun et al. \cite{sun2019know}, we leverage Part-of-Speech  (POS) tags to extract noun phrases  (NP) and verb phrases  (VP) from package description sentences as characteristics and functions of ROS packages. The regular expressions for parsing POS tags generated from the package descriptions are shown in Table \ref{table:regular-expression}.

\begin{table}[h]
\caption{Regular expressions for extracting features.}
\label{table:regular-expression}
\begin{tabular*}{\linewidth}{p{2cm}p{6cm}}
\toprule
Feature & Regular Expression \\
\midrule
Function & $ (VB.*)+ (CD)* (DT)? (CD)* (JJ)* (CD)* (VBD|VBG)* (NN.*)* (POS)* (CD)* (VBD|VBG)* (NN.*)* (VBD|VBG)* (NN.*)* (POS)* (CD)* (NN.*)+$\\
Category & $ (CD)* (DT)? (CD)* (JJ)* (CD)* (VBD|VBG)* (NN.*)* (POS)* (CD)* (VBD|VBG)* (NN.*)* (VBD|VBG)* (NN.*)* (POS)* (CD)* (NN.*)+$\\
\bottomrule
\end{tabular*}
\end{table}

Fig. \ref{fig:pos-tag} shows an example of parsing the description sentence of the ``hector\_quadrotor\_gazebo" ROS package. Each word in the sentence is tagged with a part of speech (in green color). By utilizing the previous regular expressions, we can extract the features for function and characteristics. In this example, ``provides a quadrotor model" and ``the gazebo simulator" are extracted as the function and characteristics, respectively.

With the POS tags-based method, we can retrieve both the characteristics and function features of ROS packages.

\subsection{Linking Entities}

With the extracted features, we can create various entities and relations to the RPKG. For features on robot, sensor, category, function and characteristics, they can be uniquely defined by their names. If a feature name matches an entity in the knowledge graph, we can directly link this feature to that entity. If not, we create a new entity with that feature name. For features on action, node, service, message and launch, the same feature name may refer to files with different contents. Therefore, we create a new entity for each feature name.

\subsection{Representing ROS Package Knowledge}

All current search approaches for ROS packages represent ROS package and its related features in text and search ROS package by matching text in syntax. They lack understanding and exploitation of the semantics of ROS packages, which results in their weakness when dealing with synonyms and abbreviations, and unsatisfactory performance in search accuracy. BERT is one of the most pre-trained language models in Natural Language Processing  (NLP) \cite{kenton2019bert}. It provides a representation for the semantics of text by encoding text to an embedding in high-dimensional vector spaces. For texts with similar semantics, their embeddings are also close in high-dimensional vector spaces, and vice versa. However, the BERT-Base model fails to represent robot domain terms properly into embeddings as it is adapted to general domains and its training corpus barely covers the robotics domain. For example, the word ``rplidar" is often used to refer to a type of lidar sensor. As it is not covered by the vocabulary of the BERT-Base, it will be tokenized into four word tokens, including ``r", ``\#\#pl", ``\#\#ida" and ``\#\#r", and then encoded into an embedding that does not properly represent its semantics.

\begin{table*}[ht]
    \caption{\centering{Comparisons of cosine similarity of phrase embeddings between BERT-Base and BERT-ROS}}
    \label{table:bert-sim}
    \centering
    \begin{tabular}{|c|c|c|c|}
        \hline
        \multirow{2}*{\textbf{Phrase\#1}} & \multirow{2}*{\textbf{Phrase\#2}} & \multicolumn{2}{c|}{\textbf{Cosine Similarity of Embeddings}} \\
        \cline{3-4}
        ~ & ~ & \textbf{BERT-Base}   & \textbf{BERT-ROS} \\
        \hline
        Turtlebot & Turtlebot2 & 0.834 & 0.954 \\
        \hline
        Turtlebot & Turtlebot3 & 0.831 & 0.936 \\
        \hline
        Turtlebot2 & Turtlebot3 & 0.931 & 0.915 \\
        \hline
        rplidar & rospeex & 0.796 & 0.682 \\
        \hline
        bringup rplidar	& start rplidar sensor & 0.868 & 0.885 \\
        \hline
        MoveIt framework & MoveIt library & 0.898 & 0.901 \\
        \hline
        Gazebo simulator & Gazebo simulation environment & 0.864 & 0.881 \\
        \hline
        OMPL & Open Manipulator Planning Library & 0.658 & 0.779\\
        \hline
    \end{tabular}
    
\end{table*}

We further fine-tune BERT-Base for a better embedding representation model for the ROS domain, called BERT-ROS. This involves mainly three steps:
\begin{itemize}
    \item Collect the robot context from ROS.org to build a corpus in the robot domain. 
    \item Extend the tokenizer from BERT-Base to cover the robotics domain terms.
    \item Further fine-tune BERT-Base for Masked Language Model (MLM) tasks to build BERT-ROS, which can be used to encode text with robotics domain terms into embeddings.
\end{itemize}

Table \ref{table:bert-sim} shows the cosine similarities of the embeddings for phrases with BERT-Base and BERT-ROS. The phrases ``Turtlebot" and ``Turtlebot2" are often used to refer to the second generation of the Turtlebot robot in robotic context. After fine-tuning, BERT-ROS can better distinguish them from ``Turtlebot3", the third generation of Turtlebot. ``rplidar" and ``rospeex" are two different domain terms both in syntax and in semantics so the similarity computed with BERT-ROS is lower than with BERT-Base. For synonym phrases like ``bringup rplidar" and ``start rplidar sensor", BERT-ROS also performs better. ``OMPL" is an abbreviation for ``Open Manipulator Planning Library". They are closer in embeddings representation with BERT-ROS. With improved semantic representation, the BERT-ROS model will contribute to ROS package search.

\section{RPKG-based ROS Package Search}
\label{section:search}

\begin{figure*}[!ht]
\centering
\includegraphics[width=1.0\linewidth]{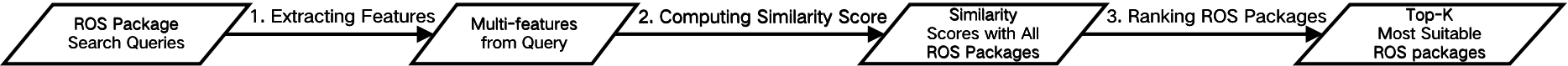}
\caption{RPKG-based ROS Package Search}
\label{fig:search}
\end{figure*}

As shown in Fig. \ref{fig:search}, there are three steps to implement the RPKG-based search for ROS packages. In the first step, feature extraction is the same as in Section \ref{section:feature}. The second step is computing similarity scores between the query and all ROS packages, which is explained in detail in Section \ref{section:similarity}. We explain the third step in Section \ref{section:algo}.

We design a fusion search algorithm that combines word-vector similarity matching and fuzzy matching to support the search of ROS packages based on RPKG. Word vectorization provides a better representation in terms of semantics, but it is not suitable for domain words in ROS packages as most of them are not common words and abbreviations are often used. On the other hand, fuzzy matching performs well in word syntactic matching, especially for domain words\cite{bo2022towards}. Hence, we combine these two methods by using vector similarity to match general functional feature words and using the fuzzy matching method to match special domain words in the descriptions of ROS packages. We rank the searched ROS packages by the weighted sum of the vector similarity score and fuzzy matching score. The algorithm is listed in Algorithm \ref{algorithm:algo}.

\begin{table}[ht]
\caption{\centering{The query template for ROS packages.}}
\label{table:template}
\begin{tabular*}{\linewidth}{p{2cm}p{6cm}}
\toprule
Features & Hints for User Input \\
\midrule
Robot & What robot do you use?\\
Sensor & What sensor do you use?\\
Category & What is the category of the package?\\
Function & What is the expected function of the package?\\
Characteristics & List the characteristics of the package, separated by commas\\
Action &  What is the action name in the package?\\
Node &  What is the node name in the package?\\
Service &  What is the service name in the package?\\
Message &  What is the message name in the package?\\
Launch &  What is the launch name in the package?\\
\bottomrule
\end{tabular*}
\end{table}

Table \ref{table:template} shows the query template we design for users to describe their requirements for ROS packages from multiple dimensions.

We formalize the ROS package $P$ as a set of features: robot $R$, sensor $SS$, category $CG$, function $F$, characteristics $CR$, action $A$, node $ND$, service $SV$, message $M$, and launch $L$. The user query $Q$ is also formalized in the same way.
$$P = \langle R, SS, CG, F, CR, A, ND, SV, M, L\rangle$$
$$Q = \langle R, SS, CG, F, CR, A, ND, SV, M, L\rangle$$

\subsection{Similarity Calculation Method}
\label{section:similarity}
For robot and sensor features, we compute the Levenshtein Distance as the similarity metric:
\begin{equation}
SIM (R_{p}, R_{q}) = \max_{i\epsilon R_{p}}(LD (i, R_{q}))
\end{equation}
For function and characteristics features, we compute the average Cosine Distance as the similarity metric:
\begin{equation}
SIM (F_{p}, F_{q}) = \frac{1}{|F_{q}|}\sum_{v_{j}\epsilon F_{q}} \max_{v_{i}\epsilon F_{p}}(\frac{v_{i}\cdot v_{j}}{|v_{i}||v_{j}|})
\end{equation}
For features in category, action, node, service, message, and launch, we compute the average of the summed Sign Function as the similarity metric:
\begin{equation}
SIM (ND_{p}, ND_{q}) = \frac{1}{|ND_{q}|}\sum_{j\epsilon ND_{q}} \sum_{j\epsilon ND_{p}} {Sgn} (i=j)
\end{equation}

We compute the weighted sum of similarities of multiple features as the similarity between each package and the query. The calculation of similarity is described in Eq. \ref{eq:sim}. $\omega_{k}$ represents the weight  ($0<\omega_{k}<=1$) for the kth feature in $P$. 
\begin{equation}
\label{eq:sim}
SIM (P, Q) = \frac{1}{\sum_{k\epsilon [0,9]} \omega_{k}}\sum_{k\epsilon [0,9]} \omega_{k} {SIM} (P_{k}, Q_{k})
\end{equation}

\subsection{Feature Fusion Search Algorithm}

We design the Feature Fusion Search Algorithm  (FFS) to search the most related ROS packages, which is shown in Algorithm \ref{algorithm:algo}.

\label{section:algo}

\begin{algorithm}[h]
\caption{Feature Fusion Search (FFS)}
\label{algorithm:algo}
\textbf{Input}: Query features: Robot $R$, Sensor $SS$, Category $CG$, Function $F$, Characteristics $CR$, Action $A$, Node $ND$, Service $SV$, Message $M$ and Launch $L$\\
\textbf{Output}: Top K most suitable packages\\
\textbf{Procedure}:
\begin{algorithmic}[1] 

    \State $Q \gets\ \langle R, SS, CG, F, CR, A, ND, SV, M, L\rangle$
    \State Initialize $S$ as an empty dictionary
    \State Initialize $R$ as an empty list
    \For{package $P$ in Package Set}
        \State $score \gets SIM (P, Q)$
        \State $S[P] \gets score$
    \EndFor
    \State $R \gets $ Sort all packages in $S$ by $score$ in descending order

    \State \textbf{return} first K packages in $R$
\end{algorithmic}
\end{algorithm}

In FFS, we first extract the features of a user query from the input. Then, we traverse all ROS packages and calculate the similarity score for each package. Finally, we sort all ROS packages by scores in descending order and output the top K packages as search results.

\section{Evaluation}
\label{section:evaluation}
The constructed RPKG has 32,294 entities  (11 types) and 54,698 relationships  (10 types), which can describe ROS packages from 10 dimensions and enrich the semantics of ROS packages. The structured knowledge representation provided by RPKG and the embedding representation of knowledge text are more convenient for supporting the subsequent search for ROS packages.

We conducted evaluations to explore the following research questions:
\begin{itemize}
    \item \textbf{RQ1  (Quality)}: How is the intrinsic quality of the knowledge in the constructed RPKG? This research question aims to evaluate the accuracy of the knowledge encapsulated within RPKG.
    \item \textbf{RQ2  (Accuracy)}: How does RPKG perform in search accuracy for ROS package compared with other methods? This research question aims to assess the accuracy of our proposed RPKG-based search method for ROS packages.
    \item \textbf{RQ3  (Necessity)}: To what extent does the removal of individual feature dimension impact the accuracy of our search method? The aim of the ablation analysis is to investigate the necessities of summarized feature dimensions for ROS package search.
\end{itemize}

\subsection{RQ1: Quality of the Constructed RPKG}
Since knowledge of category, node, service, message, and launch is extracted from the structured code file system, it is inherently accurate. Our quality evaluation focuses on the robot, sensor, function, and characteristics related to the ROS package.

\subsubsection{Experiment Setup}
Similar to previous studies \cite{li2018improving,ren2020api,liu2020generating,bo2022towards}, we adopt a sampling method \cite{singh2013elements} to ensure that the ratios observed in the sample are representative of the population within a 95\% confidence level, with a confidence interval of 5.

We randomly select 96 hardware devices, 344 functions, and 371 characteristics to conduct the experiment. Two developers  (who are not involved in this study and are familiar with ROS) independently perform the examination. Given a ROS package and a feature in each round, developers are required to determine whether this feature is related to the package in a specific dimension. All decisions are binary  (the accuracy rates are Accuracy\#1 and Accuracy\#2 respectively). For the data instances where the two annotators disagree, they have to discuss and come to a consensus. We compute the final accuracy after resolving the disagreements  (Accuracy\#F) and compute Cohen’s Kappa agreement (Kappa Agreement) \cite{landis1977application} to evaluate the inter-rater agreement. Based on the consensus annotations, we evaluate the quality of the knowledge created about the ROS package.

\subsubsection{Results and Analysis}

\begin{table*}[ht]
    \caption{\centering{Accuracy of ROS Package Features}}
    \label{table:acc}
    \centering
    \begin{tabular}{|c|c|c|c|c|}
        \hline
        \textbf{Aspect} & \textbf{Accuracy\#1}  & \textbf{Accuracy\#2} & \textbf{Accuracy\#F} & \textbf{Kappa Agreement} \\
        \hline
        Hardware  & 95.8\%   & 96.9\%   & 95.8\%  &  0.852  \\
        \hline
        Function & 93.0\% & 90.1\% & 91.6\% & 0.898   \\
        \hline
        Characteristics & 90.3\% & 92.5\% & 93.0\% & 0.748\\
        \hline
    \end{tabular}
    
\end{table*}

The results are shown in Table \ref{table:acc}. We can see the agreement rate are all above 0.74, indicating substantial or almost perfect agreement between the two annotators. The accuracy is generally high  (above 0.91), which represents the high quality of our RPKG.

Typical problems of ROS package hardware extraction include:  (1) mismatched semantics, for example, the word ``point" in ``point\_cloud\_publisher\_tutorial" was incorrectly matched to the ``pointgrey camera". (2) Incomplete hardware list, for example, the package ``baxter\_gazebo" is developed for the Baxter robot. However, this robot is not included in the official ROS robot list, so the fuzzy match-based method will miss this feature.

Typical problems of ROS package function and characteristics extraction include:  (1) POS tagging error, for the ``audio\_play" package with the description ``Outputs audio to a speaker from a source node.", ``Outputs" is tagged as a noun and the phrase ``Outputs audio" is recognized as a characteristics feature, when in fact it is a function feature.  (2) Meaningless characteristics, ``This package" is extracted as a characteristics feature from ``This package provides launch files for operating Care-O-bot.".  (3) Morphological difference, the noun phrase ``face detection" from ``Face detection in images." describes the function ``detect face", but it is extracted as a characteristics feature.

\begin{mdframed}[backgroundcolor=gray!20]
The results also prove that our proposed multi-dimensional feature extraction technique is effective and can support the knowledge extraction for ROS packages.
\end{mdframed}

\subsection{RQ2: Accuracy of the Search Method}

\label{section:rq2}
To validate the enhanced accuracy of ROS package searching, we conducted comparative case studies, comparing our semantic-based search approach with four baseline search approaches: ROS Index, GitHub, Google, and ChatGPT. Through these case studies, we gained valuable insights into the strengths and limitations of our approach.

\subsubsection{Experiment Setup}
To evaluate the effectiveness of our search approach, we selected 29 robot tasks from three ROS tutorial books (Mahtani, 2016 \cite{mahtani2016effective}; Fairchild, 2017 \cite{fairchild2017ros}; Gandhinathan, 2019 \cite{gandhinathan2019ros}), which cover common tasks in robotics, such as navigation and manipulation. These three referenced books provide rich ROS-based robot software development cases, aiming to guide newcomers to quickly get started with ROS.

As shown in Table \ref{table:query-task}, each task involves several subtasks. Each subtask requires a ROS package. We summarized 100 subtasks, involving 100 ROS packages. According to the template shown above in Table \ref{table:template}, we formulated a query based on the description for each subtask. For example, subtask\#4-3 requires to ``visualize Turtlebot2 with RViz", whi-ch was formulated into the query with ``$\langle$ Robot:Tu-rtlebot2, Function:visualize Turtlebot2, Characteri-stics:RViz$\rangle$". We know that the desired ROS package is ``turtlebot\_rviz\_launchers".

For features on function and characteristics, we assigned a weight of 0.8 to achieve better search performance, while we set it to 1.0 for other feature dimensions. We set 5 search levels, with K values of 1, 5, 10, 15, and 20.

\subsubsection{Baselines}
We use ROS Index, GitHub, Google, and ChatGPT as the baseline tools. Each of them implements a different search algorithm.

\begin{itemize}
    \item \textbf{ROS Index}: It uses the BM25 algorithm to calculate the relevance score between ROS packages and user queries. Designed to search ROS packages and repositories, ROS Index is the most commonly used search platform, especially for ROS newcomers.
    \item \textbf{GitHub}: It uses Elasticsearch as its search engine, which implements the inverted index algorithm to search data. Moreover, most of the official ROS packages are hosted on GitHub.
    \item \textbf{Google}: It uses PageRank and many other text match-based algorithms to search for relevant documents. To improve the search results, we can use the ``site" keyword to limit its search scope to include only ``ros.org" and ``github.com".
    \item \textbf{ChatGPT}: It uses large-scale language models, including gpt-3.5-turbo, which greatly improves the ability to understand the semantics of texts and codes.
\end{itemize}

The inputs differ in format between the four baselines and our method. For example, if the search input consists of more than 3 words, then GitHub may not return any results. Hence, we determine the best input for each task through several trials, which helps us find the desired ROS packages.

\subsubsection{Results and Analysis}

\begin{figure*}[h]
\centering
\includegraphics[width=0.8\linewidth]{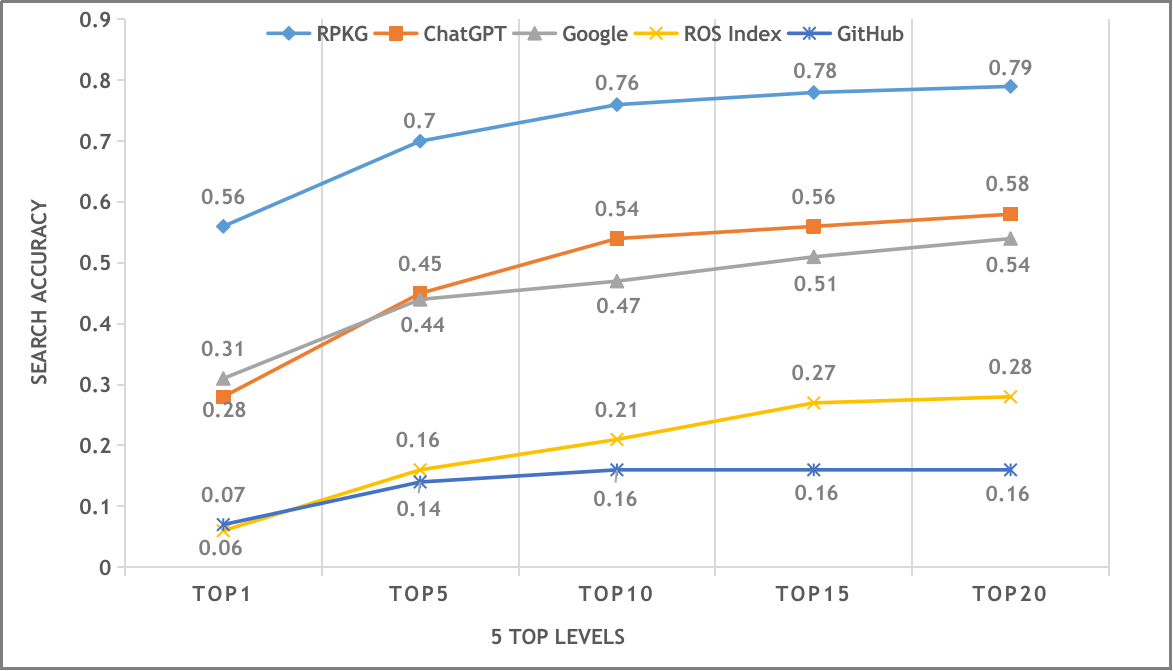}
\caption{Comparison of search accuracy on 5 top levels}
\label{fig:accuracy}
\end{figure*}

Fig. \ref{fig:accuracy} shows the comparisons on search accuracy between the existing four methods and our RPKG-based search method. For each query, if the desired ROS package appears in the top@K search results, then this set of K ROS packages is considered accurate. From this table, we can observe a significant improvement of up to 21\%, which demonstrates that our method can find more accurate ROS packages. ChatGPT exhibits slightly higher accuracy than Google at four top levels due to its better semantic understanding.

\afterpage{%
  \clearpage 

\begin{table*}[ht]
    \caption{\centering{Statistics of 6 tasks including 20 subtasks from reference books}}
    \label{table:query-task}
    \centering
    \begin{tabular}{p{1.5cm}p{6cm}p{5.5cm}p{3cm}}
        \hline
        \textbf{Task No.} & \textbf{Subtask Description} & \textbf{Query for ROS Package} & \textbf{Desired Package} \\
        \hline
        \multirow{3}*{1} & 1-1 Discover other ROS masters &  \textbf{Function}:discover other ROS masters & master\_discovery\_fkie \\
        \cline{2-4}
        ~ & 1-2 Synchronize the local ROS master & \textbf{Function}:synchronize the local ROS master & master\_sync\_fkie \\
        \cline{2-4}
        ~ & \makecell[l]{1-3 Use GUI to pilot the robot with Twist \\message} & \makecell[l]{\textbf{Function}:pilot the robot\\\textbf{Characteristics}:GUI, Twist message} & rqt\_robot\_steering \\
        
        \hline

        \multirow{3}*{2} & \makecell[l]{2-1 Meta-package for easily installing \\Velodyne simulation components} &  \makecell[l]{\textbf{Sensor}:Velodyne HDL-64E 3D lidar\\\textbf{Function}:install velodyne \\\textbf{Category}:meta package} & velodyne\_simulator \\
        
        \cline{2-4}
        ~ & 2-2 Description package for Velodyne lidar & \makecell[l]{\textbf{Sensor}:Velodyne HDL-64E 3D lidar\\\textbf{Category}:description package} & velodyne\_description \\
        
        \cline{2-4}
        ~ & \makecell[l]{2-3 Use Gazebo plugin to provide simulated  \\data from Velodyne lidar} & \makecell[l]{\textbf{Sensor}:Velodyne HDL-64E 3D lidar\\\textbf{Function}:provide simulated data\\\textbf{Characteristics}:Gazebo plugin} & velodyne\_gazebo\_plugins \\
        
        \hline

        \multirow{2}*{3} & \makecell[l]{3-1 Simulate Turtlebot2 in Gazebo} &  \makecell[l]{\textbf{Robot}:Turtlebot2\\\textbf{Function}:simulate Turtlebot2\\\textbf{Characteristics}:Gazebo} & turtlebot\_gazebo \\
        
        \cline{2-4}
        ~ & \makecell[l]{3-2 Open dashboard for Turtlebot2} & \makecell[l]{\textbf{Robot}:Turtlebot2\\\textbf{Function}:open dashboard} & turtlebot\_dashboard \\
        \hline

        \multirow{5}*{4} & \makecell[l]{4-1 Start the Turtlebot2 robot} & \makecell[l]{\textbf{Robot}:Turtlebot2\\\textbf{Function}:start the TurtleBot2} & turtlebot\_bringup \\

        \cline{2-4}
        ~ & \makecell[l]{4-2 Use Turtlebot2 to create maps with \\gmapping\_demo.launch} & \makecell[l]{\textbf{Robot}:Turtlebot2\\\textbf{Function}:create maps\\\textbf{Launch}:gmapping\_demo.launch} & turtlebot\_navigation \\

        \cline{2-4}
        ~ & \makecell[l]{4-3 Visualize Turtlebot2 with RViz} & \makecell[l]{\textbf{Robot}:Turtlebot2\\\textbf{Function}:visualize Turtlebot2\\\textbf{Characteristics}:RViz} & turtlebot\_rviz\_launchers \\
        
        \cline{2-4}
        ~ & \makecell[l]{4-4 Teleoperate Turtlebot2 with joysticks or \\keyboard} & \makecell[l]{\textbf{Robot}:Turtlebot2\\\textbf{Function}:teleoperate Turtlebot2\\\textbf{Characteristics}:joysticks, keyboard} & turtlebot\_teleop \\

        \cline{2-4}
        ~ & \makecell[l]{4-5 Save the map} & \makecell[l]{\textbf{Function}:save the map} & map\_server \\

        \hline
        \multirow{4}*{5} & \makecell[l]{5-1 Package that includes tools for Bebop \\robot} & \makecell[l]{\textbf{Function}:include tools\\\textbf{Characteristics}:Bebop} & bebop\_tools \\

        \cline{2-4}
        ~ & \makecell[l]{5-2 Message package for Bebop robot} & \makecell[l]{\textbf{Characteristics}:Bebop\\\textbf{Category}:message package} & bebop\_msgs \\

        \cline{2-4}
        ~ & \makecell[l]{5-3 Package that includes \\bebop\_driver\_node node} & \makecell[l]{\textbf{Characteristics}:Bebop\\\textbf{Node}:bebop\_driver\_node} & bebop\_driver \\
        
        \cline{2-4}
        ~ & \makecell[l]{5-4 Description package for Bebop robot} & \makecell[l]{\textbf{Characteristics}:Bebop\\\textbf{Category}:description package} & bebop\_description \\
        
        \hline

        \multirow{3}*{6} & \makecell[l]{6-1 Start the setup helper for MoveIt} & \makecell[l]{\textbf{Function}:start the setup helper\\\textbf{Characteristics}:MoveIt} & moveit\_setup\_assistant \\

        \cline{2-4}
        ~ & \makecell[l]{6-2 Provide base 3D markers} & \makecell[l]{\textbf{Function}:provide base markers\\\textbf{Characteristics}:3D marker}  & aruco\_ros \\

        \cline{2-4}
        ~ & \makecell[l]{6-3 Start the server for grasp with MoveIt} & \makecell[l]{\textbf{Function}:start the server\\\textbf{Characteristics}:grasp, MoveIt}  & moveit\_simple\_grasps \\

        \hline

    \end{tabular}
\end{table*}
\clearpage 
}

We analyze those query cases and find the reason why our method performs better is that we greatly improve the representation and understanding of ROS packages by extracting richer semantic features. For those cases where our method performs poorly, to sum up, the first reason is that the pre-trained language model we fine-tuned, BERT-ROS, has limited semantic representation and understanding capabilities, and its understanding of user query intentions is not as good as that of ChatGPT. The second reason is that certain ROS software packages have not been given relevant descriptions by developers, which results in few features extracted. In the next step, we plan to combine large language models such as GPT3.5 represented by ChatGPT and knowledge graphs to enhance semantic representation and understanding for ROS package search task.

\begin{table}[h]
    \caption{Comparison of search accuracy on different sample sizes}
    \label{table:solid}
    \centering
    \begin{tabular}{|c|c|c|c|c|c|}
        \hline
        \textbf{Sample Size} & \textbf{Top1} & \textbf{Top5} & \textbf{Top10} & \textbf{Top15} & \textbf{Top20} \\
        \hline
10 & 0.70 & 	0.80 & 	0.80 & 	0.90 & 	0.90 \\
        \hline
20 & 0.65 & 	0.70 & 	0.75 & 	0.75 & 	0.75\\
        \hline
30 & 0.50 & 	0.73 & 	0.73 & 	0.80 & 	0.80\\
        \hline
40 & 0.63 & 	0.73 & 	0.78 & 	0.80 & 	0.80\\
        \hline
50 & 0.58 & 	0.72 & 	0.80 & 	0.80 & 	0.82\\
        \hline
60 & 0.55 & 	0.73 & 	0.80 & 	0.82 & 	0.82\\
        \hline
70 & 0.56 & 	0.71 & 	0.77 & 	0.79 & 	0.80\\
        \hline
80 & 0.58 & 	0.73 & 	0.76 & 	0.78 & 	0.79\\
        \hline
90 & 0.60 & 	0.72 & 	0.79 & 	0.81 & 	0.82\\
        \hline
100 & 0.56 & 	0.70 & 	0.76 & 	0.78 & 	0.79\\
        \hline

    \end{tabular}
    
\end{table}

Table \ref{table:solid} shows the comparisons of search accuracy on different sample sizes. We can find that the variations in search accuracy range between 0.1 and 0.2 when the sample size is no more than 30. When the sample size exceeds 30, the variations in search accuracy are within 0.1 across 5 top levels, which indicates the robustness and consistency of our method. This consistent performance suggests that our method's search accuracy is relatively stable even when confronted with varying sample sizes. This finding reinforces our method's reliability and reinforces our confidence in its suitability for practical applications.

\begin{figure}[H]
\begin{mdframed}[backgroundcolor=gray!20]
We conducted extensive experiments to evaluate the effectiveness of our proposed ROS package search approach. The results clearly demonstrate that our approach significantly outperforms existing methods in terms of search accuracy and relevance.
\end{mdframed}
\end{figure}

\subsection{RQ3: Necessity of Multi-dimensional Features}

Five feature dimensions  (robot, sensor, category, function, and characteristics) are summarized for searching ROS packages in section \ref{section:rpkg-construction}. In order to verify their necessity for our search method, we set up five individual ablation experiments to observe their impacts on the search accuracy for ROS packages. The code file features, including action, node, service, message, and launch, are inherent structural features of a ROS package, so there is no need to carry out ablation experiments for them. The ablation analysis serves to elucidate the significance of each feature dimension within our search method. This investigation enables us to quantify the impact of individual features on the accuracy of package retrieval, thereby providing a clearer understanding of the role that each dimension plays in enhancing the search effectiveness of our approach.

\subsubsection{Experiment Setup}
We select five dimensions of features: robot, sensor, category, function, and characteristics. For each dimension, we construct two sets of queries: one with the inclusion of the specific feature and another with the exclusion of that feature. From a pool of 100 diverse query tasks in section \ref{section:rq2}, we identified queries that contain each specific feature. We identified 15 queries with the ``robot" feature, 16 queries with the ``sensor" feature, 14 queries with the ``category" feature, 52 queries with the ``function" feature, and 82 queries with the ``characteristics" feature.

For each of the five feature dimensions, we created two distinct query groups:
\begin{itemize}
    \item \textbf{Inclusion Group}: This group comprises queries that include the feature under consideration. For instance, the ``robot inclusion" group consists of queries containing the ``robot" feature.
    \item \textbf{Exclusion Group}: In contrast, the exclusion group contains queries where the specific feature is deliberately removed.
\end{itemize}

\subsubsection{Results and Analysis}

For each query group within every feature dimension, we conduct a series of searches using our method. We record the accuracy of search results for each query task in terms of whether the relevant ROS package is successfully retrieved on 5 top levels.

\begin{figure*}[h]
    \centering 
    
    \subfigure[``robot" feature]{
        \centering 
        \includegraphics[width=0.3\linewidth, height=3.6cm]{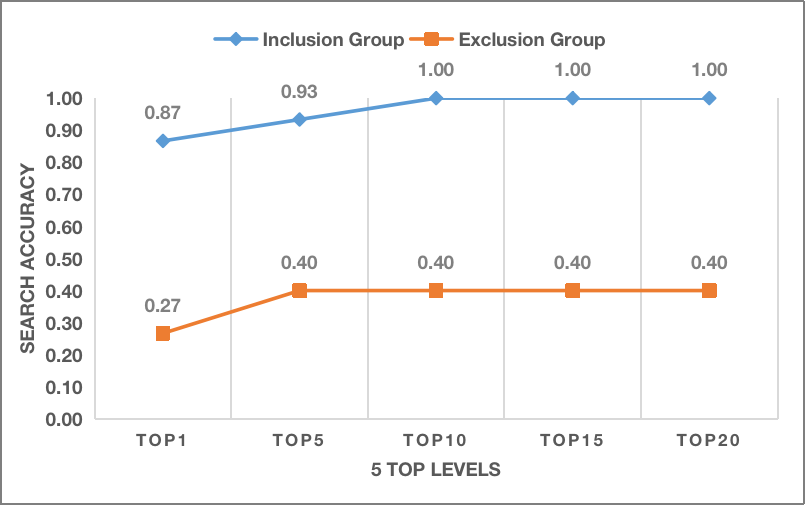} 
        \label{fig:necessity-robot}}
    \subfigure[``sensor" feature]{
        \centering
        \includegraphics[width=0.3\linewidth, height=3.6cm]{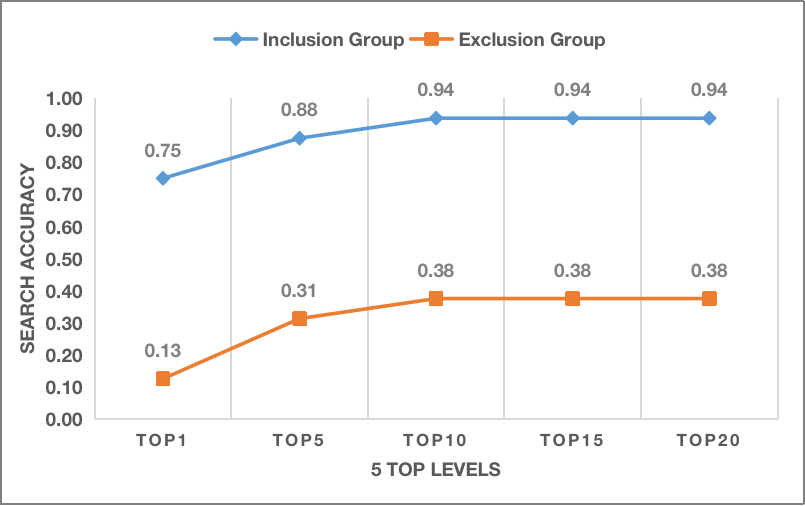}
        \label{fig:necessity-sensor}}
    \subfigure[``category" feature]{
        \centering
        \includegraphics[width=0.3\linewidth, height=3.6cm]{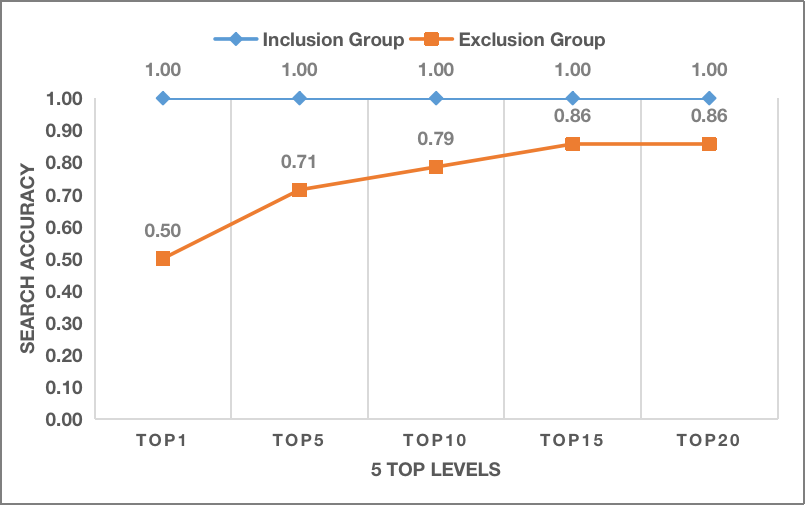}
        \label{fig:necessity-category}}
    
    \subfigure[``function" feature]{
        \centering
        \includegraphics[width=0.3\linewidth, height=3.6cm]{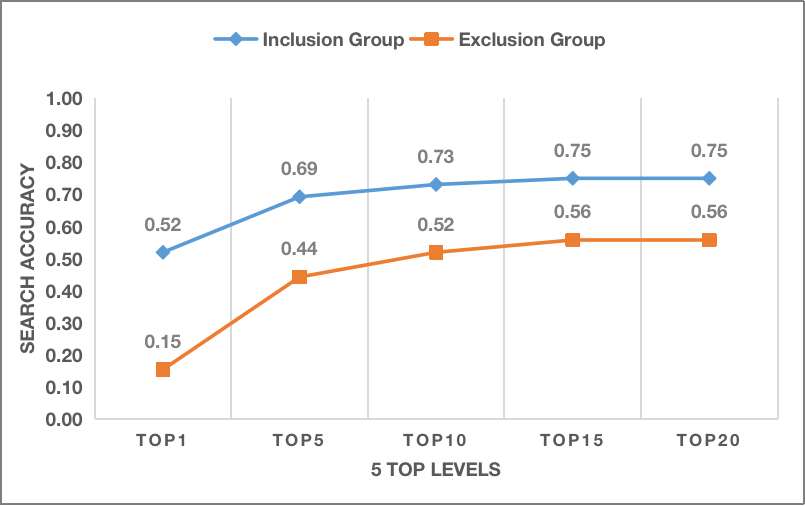}
        \label{fig:necessity-function}}
    \subfigure[``characteristics" feature]{
        \centering
        \includegraphics[width=0.3\linewidth, height=3.6cm]{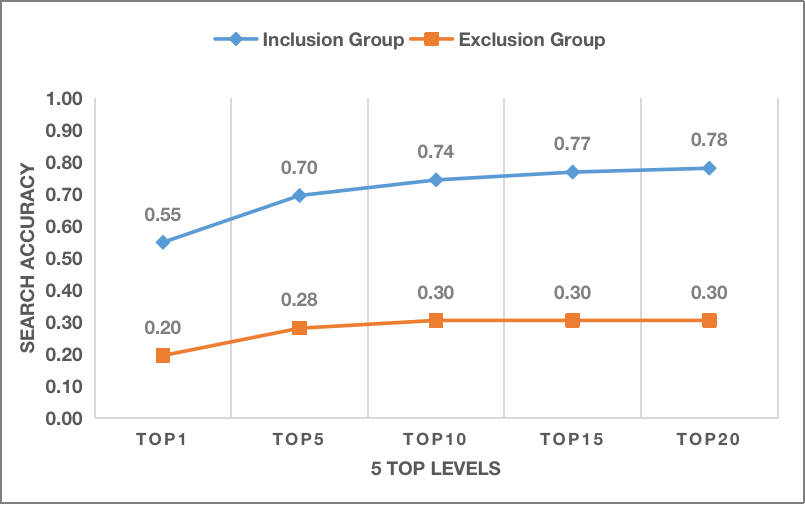}
        \label{fig:necessity-characteristics}}

    \caption{Comparisons on search accuracy between inclusion group and exclusion group for five feature dimensions}
    \label{fig:necessity}

\end{figure*}

Fig. \ref{fig:necessity} presents the accuracy variation of our search method before and after ablation for each of the five feature dimensions: robot, sensor, category, function, and characteristics. From the five sub-figures, it is evident that the exclusion groups exhibit a noticeable decrease in search accuracy. Furthermore, it can be observed that the features ``robot" and ``sensor" have a more pronounced impact on the search method compared to other features. This suggests that these two features play a more crucial role in enhancing the effectiveness of ROS package search.

\begin{figure}[H]
\begin{mdframed}[backgroundcolor=gray!20]
We conducted ablation experiments to evaluate the necessity of our summarized five feature dimensions. The results clearly reflect the importance of the features we summarize, and their different degrees of influence on the search accuracy.
\end{mdframed}
\end{figure}

\section{Discussion}
\label{section:discussion}

\subsection{Implications}

For researchers, they can build upon this work by further mining features of ROS packages from code files or optimizing search method. For developers, they can query related knowledge of certain ROS packages from RPKG or search related ROS packages for their robot software projects. For tool makers, this work can guide the design of new search tools for ROS packages. 

\subsection{Threats to Validity}

A threat to internal validity is the subjective judgment in evaluation for the quality of extracted knowledge. To alleviate this threat, we have provided the agreement level for all subjective judgments. Another threat to internal validity is that our dictionary for identifying hardware features of ROS packages may not cover all the latest devices in robotics, including many moving bases, sensors, and UAVs. In the future, we plan to launch an open project for collecting robotic hardware information and domain-specific words about algorithms, tools, etc., which will contribute to reducing this threat.

The major threat to external validity is the generalization of our experiment for our search method. To alleviate this threat, we are going to apply our approach to more open tasks related to searching ROS packages in Q\&A platforms like ROS Answers and release our knowledge graph for public evaluations.

\section{Conclusion and Future Work}
\label{section:conclusion}
In conclusion, this research article aims to address the problem of searching for ROS packages by pro-posing a comprehensive approach that integrates feature extraction, knowledge graph representation, semantic matching, and a search method based on weighted similarities.

Firstly, we introduce a novel feature extraction technique specifically designed for ROS packages. This technique allows us to extract multi-dimensional features that capture the essential characteristics of the packages, enabling a more accurate and effective search.

Secondly, we construct a knowledge graph to store and represent these extracted features as knowledge. The knowledge graph provides a structured representation of the relationships between different ROS packages, facilitating the retrieval of relevant information and enhancing the search process.

Furthermore, to support semantic matching, we fine-tune a BERT model specifically tailored to the ROS domain. This model allows us to encode the textual information associated with ROS packages into meaningful embeddings, enhancing the understanding and similarity computation between packages.

Finally, we design a search method that leverages the computed sum of weighted similarities across multiple features. By assigning appropriate weights to different features, we can prioritize and find the most related ROS packages to users, ensuring that the found ROS packages are both relevant and useful.

Overall, our approach offers a comprehensive solution to the problem of searching for ROS packages, integrating feature extraction, knowledge representation, semantic matching, and a robust search method. Through our experiments and evaluations, we demonstrate the quality and accuracy of our proposed approach, showcasing its potential to enhance the discovery and utilization of ROS packages in various robotic applications.

As a next step, we intend to further study how to mine software design patterns from existing ROS-based robot software in ROS wiki tutorials, instructional books, and research papers. We also plan to further study how to utilize the mined design patterns to support the development of robot software.

\begin{acknowledgement}
This work was supported by the National Natural Science Foundation of China  (Grant Nos. 62172426).
\end{acknowledgement}

\bibliographystyle{fcs}
\bibliography{ref}

\begin{biography}{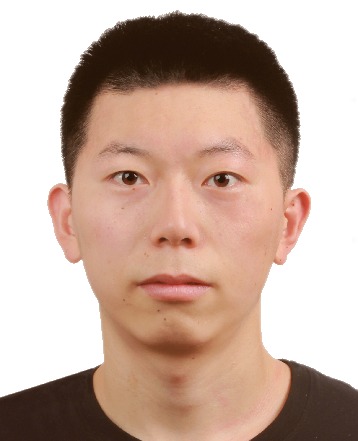}
Shuo Wang is a Ph.D graduate student in the College of Computer, National University of Defense Technology, China.  His research interests include robotics software engineering, data mining and knowledge graph.
\end{biography}

\begin{biography}{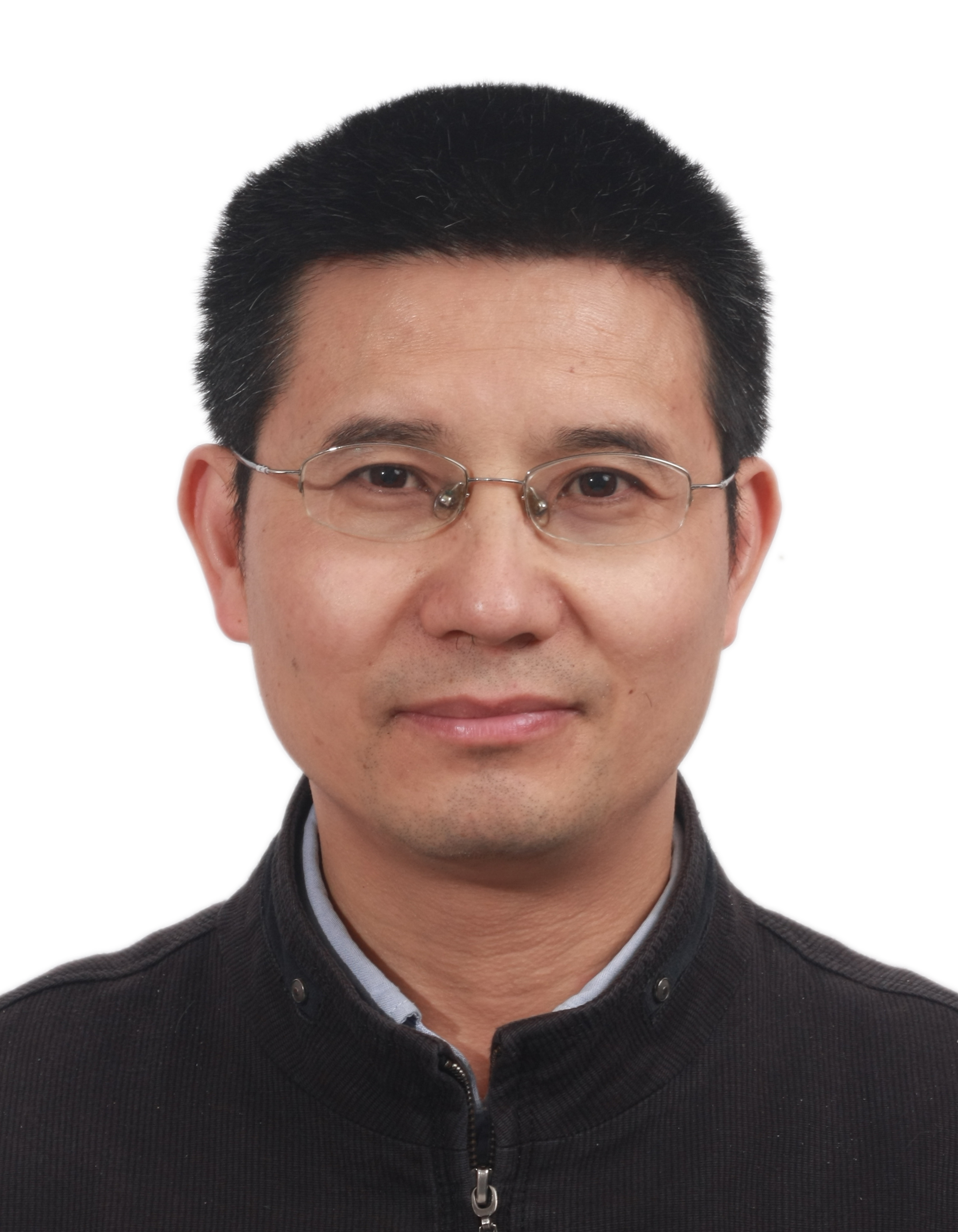}
Xinjun Mao is a professor in the College of Computer, National University of Defense Technology, China. He received his Ph.D. degree in computer science from the National University of Defense Technology, China in 1998. His research interests include intelligent software engineering, multi-agent system, autonomous robot software, self-adaptive system, and crowdsourcing.
\end{biography}

\begin{biography}{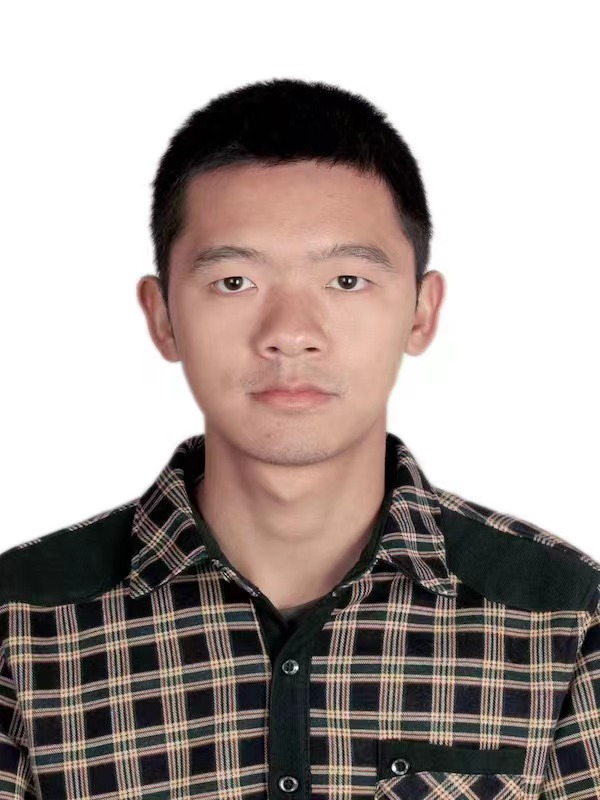}
Shuo Yang is a lecturer in the College of Systems Engineering, National University of Defense Technology, China. He received his Ph.D. degree in software engineering from the National University of Defense Technology, China in 2022. His research interests include robotics software engineering, automated planning, and component-based robotic frameworks.
\end{biography}

\begin{biography}{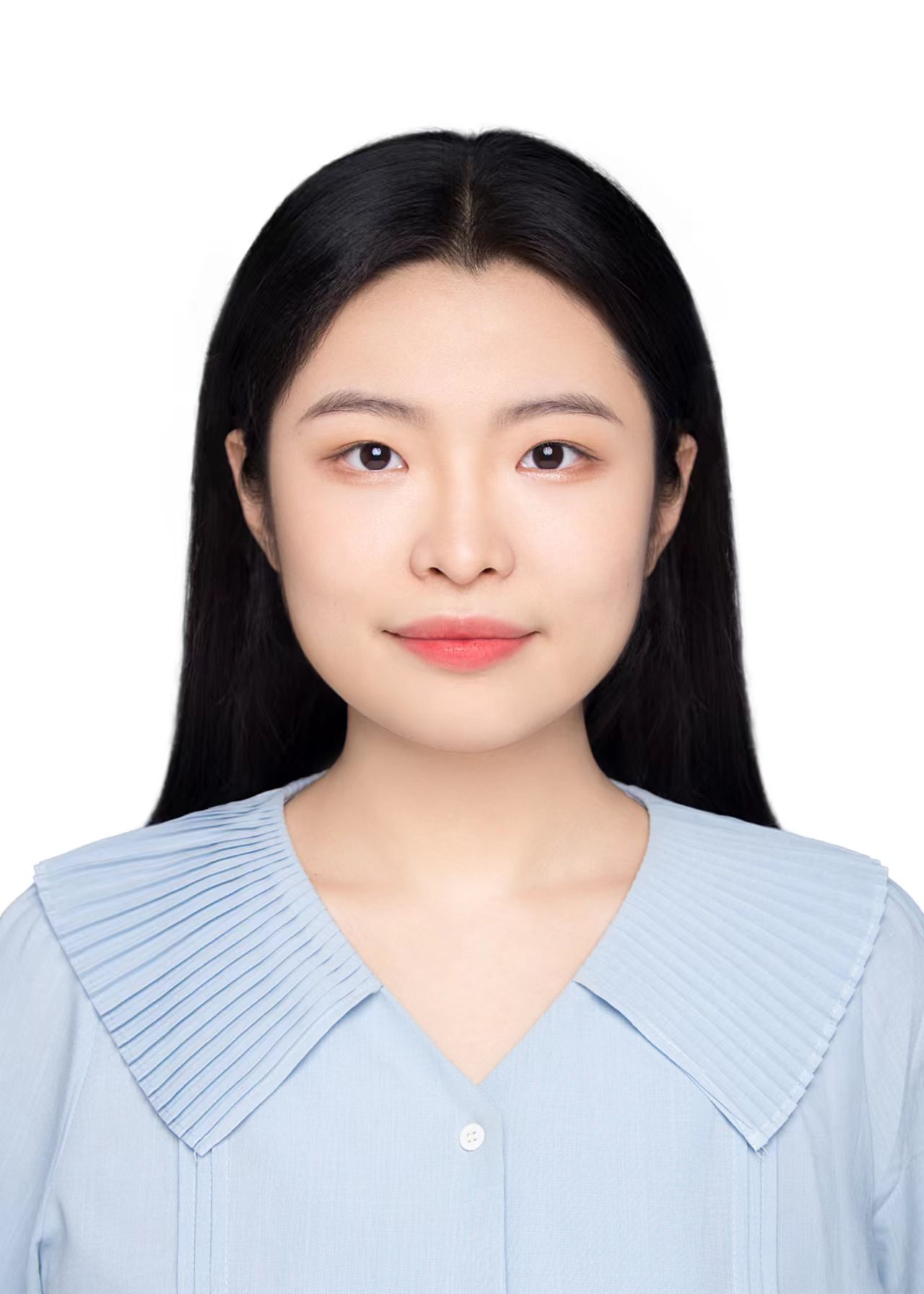}
Menghan Wu is a Ph.D. student at the College of Computer Science and Technology, Zhejiang University, China. Her  current  research  interests  include artificial intelligence for software engineering.
\end{biography}

\begin{biography}{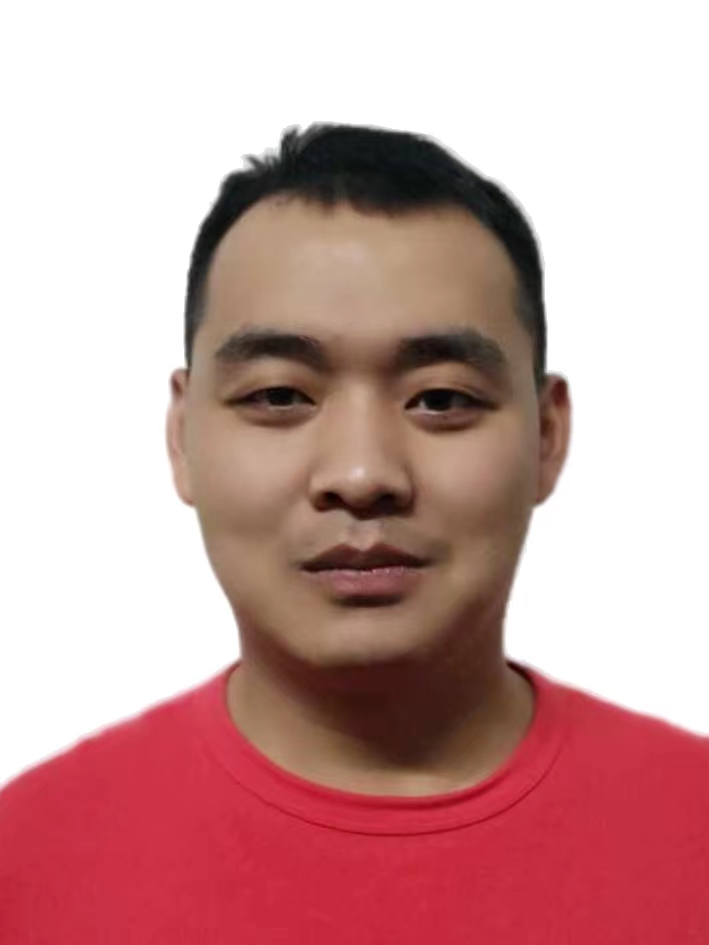}
Zhang Zhang is a Ph.D graduate student in the College of Computer, National University of Defense Technology, China. His work interests include open source software engineering, data mining, and crowdsourced learning.
\end{biography}

\end{document}